%% file: Slonimczyk_Kimsanbaev_arxiv_v0.tex
\documentclass[12pt]{article}
\usepackage[margin=1in]{geometry}
\usepackage{amsmath,amssymb}
\usepackage{graphicx}
\graphicspath{{figures/}}
\usepackage{booktabs}
\usepackage{makecell}
\usepackage{multirow}
\usepackage{array}
\usepackage{colortbl}
\usepackage{tikz}
\usepackage[most]{tcolorbox}
\usepackage{ctable}
\usepackage{natbib}
\bibpunct{(}{)}{;}{a}{,}{,}
\usepackage{setspace}
\usepackage{caption}
\usepackage{subcaption}
\usepackage{hyperref}
\usepackage{xcolor}
\usepackage{epigraph}
\usepackage{pifont}

\newcommand{\RVE}{\mathrm{RVE}}

\newcommand{\ballotcheck}{\text{\ding{51}}} 
\newcommand{\ballotx}{\text{\ding{55}}}     
\newcommand{\chkmrk}{$\textcolor{green}{\ballotcheck}$}
\newcommand{\xmrk}{$\textcolor{red}{\ballotx}$}

\title{Good Looks and the Academic Job Market: Measuring Attractiveness with Vision Embeddings}
\author{Fabi\'an Slonimczyk\thanks{ICEF, Higher School of Economics. Contact: \texttt{fslonimczyk@hse.ru}} \and
        Shergazy Kimsanbaev\thanks{ICEF, Higher School of Economics.}}
\date{This version: November 2026}

\begin{document}

\maketitle

\begin{abstract}
\noindent We introduce a new approach to measure physical attractiveness---the \emph{relative vision embedding} (RVE) score. The score maps a face to a vector in the embedding space of a pre-trained deep vision model and measures its proximity to a reference set of celebrity faces. It requires no additional training or fine-tuning, no hand-crafted features, and no human raters; it is continuous, inexpensive, scalable, and fully reproducible. Using publicly available
photographs with human attractiveness ratings, we show that the RVE scores correlate strongly with human ratings and that rankings are consistent across several off-the-shelf vision models. We apply the measure to administrative hiring data from a research-oriented economics department, complemented by hand-collected photographs of the job applicants. Higher RVE scores are
associated with better job-market outcomes, though the association is not robust
across all specifications. In contrast, the association with reference-letter length and sentiment is strong and robust: candidates with higher RVE scores receive longer letters written in more favorable language, and this sentiment gradient is steeper for female candidates. The pattern is consistent with appearance affecting academic hiring in part through letters of recommendation.\\

\vspace{0.5em}
\end{abstract}

\onehalfspacing

\newpage

\epigraph{\dots the handsome person can attain all his ends without doing anything.}%
         {Xenophon, \emph{Symposium}}
\vspace{0.5em}

\section{Introduction}

Physical attractiveness is a scarce and valued attribute. The beauty premium is well documented: attractive workers earn more than otherwise observationally similar colleagues, whether in cross-sectional data \citep{Hamermesh_Biddle_1994,Biddle_Hamermesh_1998}, over the life course
\citep{Scholz_Sicinski_2015}, or in a laboratory labor market
\citep{Mobius_Rosenblat_2006}. Correspondence studies in Sweden, Israel, Argentina, and China all find that attractive applicants are
more likely to be invited to interview
\citep{Rooth_2009,Ruffle_Shtudiner_2015,LopezBoo_et_al_2013,Deng_Li_Zhou_2020}. The premium extends beyond the labor market
to life satisfaction \citep{Hamermesh_Abrevaya_2013}, electoral success
\citep{Berggren_et_al_2010}, and early cognitive development \citep{Hamermesh_et_al_2023},
among other outcomes.\footnote{\citet{Mocan_Tekin_2010} show that less attractive individuals are
more likely to commit several types of crimes, with a non-trivial part of the effect mediated
by labor market opportunities.}

With few exceptions, the literature relies on human raters to measure physical attractiveness. This approach has several limitations. First, ratings can be highly variable. Raters disagree with one another, so the same face must be evaluated by several raters to obtain a reliable measure. This is costly and limits how many faces a study can rate. Second, beauty standards vary across cultures and over time, so ratings collected in different settings are not directly comparable. Third, judgments may be colored by ascriptive characteristics of the person being rated, such as race, gender, or age. This makes it difficult to separate the returns to appearance from the returns to those characteristics. Fourth, human ratings are typically coarse. Most studies rely on ordinal scales with only a handful of categories. Such scales discard much of the variation in a fundamentally continuous trait and attenuate estimated effects. Finally, the judgments behind any set of scores cannot be recovered, by the original authors or by anyone else. In short, human ratings constrain what attractiveness studies can measure, how precisely they can measure it, how large those studies can be, and whether their findings can be reproduced.

This paper proposes a new measure that addresses these limitations. Our approach rests on three observations. The first is that pre-trained vision models encode the features present in an image as a high-dimensional vector called an embedding, and the geometry of this embedding space preserves visual similarity. Mapping a face into this space therefore yields a representation that reflects its visual appearance. The second is that physical attractiveness is a relative quality, so it is naturally measured by comparison rather than on an absolute scale. The third is that, although individual ratings are noisy, aggregate judgments of attractiveness nonetheless exhibit broad agreement. Some individuals are widely regarded as highly attractive, and they can serve as a natural reference set. Together, these observations suggest a simple construction: represent each face as a vector and measure its physical attractiveness as the average proximity to the reference set in the embedding space. We refer to this measure as the \emph{Relative Vision Embedding} (RVE) score.

Several properties follow from this construction. Because nothing is estimated or tuned on the sample being scored, the value assigned to an image does not depend on the other images in the study. A face receives the same score whether it is scored alone or alongside many others. Human ratings lack this context-independence, since raters judge each face relative to other faces and the rating scale presented to them. The score is also inexpensive to produce. Once the model and the reference embeddings are in place, scoring an image requires a single forward pass, so the marginal cost is negligible and the measure scales to sample sizes far larger than any rater panel could handle.

Reproducibility follows from the same facts. The model, the reference set, and the metric are the only inputs, and all three can be documented and released. Anyone who uses these inputs obtains identical scores for identical images, and scores remain comparable across place and time as long as the reference set is held constant. The score is continuous, so it preserves variation that ordinal scales discard and can enter linear models without the information loss caused by discretization.

The baseline reference set is drawn from \emph{Ranker}, a public ranking website where visitors vote on lists of celebrities ranked by physical attractiveness. Its rankings aggregate hundreds of thousands of votes cast by millions of visitors, so the ordering reflects a broad sample of public opinion. From these rankings we construct separate lists for each race and gender, with up to 100 celebrities per list, covering Asian, Black, Latino, and White celebrities.\footnote{\url{https://www.ranker.com}, accessed February 7th 2026. Appendix A reproduces a few images from each list. The eight lists, their URLs, the reference photographs, and PDF snapshots of each webpage are available at \url{https://github.com/fabianslo108/[TBC]}.}

We conduct a few validation exercises to verify that the RVE is aligned with human judgement. First, we use the Chicago Face Database 3.0 (CFD), a publicly available resource consisting of photographs of 597 individuals, each rated for attractiveness by 28 independent volunteers, together with self-reported gender and race. We show that the RVE score correlates strongly
with the human ratings, especially when controlling for gender and race. We also examine the geometry of the embedding space
directly: a principal-components projection and a $k$-means clustering of the pooled
celebrity and CFD images recover clusters that align with gender and with the source of the photograph. Specifically, the modal class of each cluster is, respectively, CFD males (94.7\%), CFD females (98\%), celebrity males (99.7\%), and celebrity females (92.8\%). This last result is reassuring, as it confirms that the embedding space carries the information we need. Finally, we use a leave-one-out procedure to directly compare RVE scores for celebrities and CFD individuals. For all races and genders, there is a sizeable gap between the two groups.

We apply the measure to the academic economics job market. Using administrative hiring data from a research-oriented economics department in Russia between 2013 and 2021, we study the relationship between physical attractiveness, reference letters, and job market outcomes. The setting is well suited to this question because the applicant pool is relatively homogeneous in training and seniority, the hiring process is highly standardized, and confidential reference letters are available. We hand-collected at least one photograph from online sources for each of 630 job-market applicants.

We report two findings. First, higher RVE scores are associated with better job-market outcomes, but the association is weak and not robust. Second, higher RVE scores are associated with systematically stronger reference letters. Better-looking candidates receive longer letters and letters written in more favorable language, and these associations survive an extensive set of controls for demographics, Ph.D. program, and letter-writer characteristics. The sentiment gradient is steeper for female candidates than for male candidates. Taken together, the two findings are consistent with a specific mechanism: appearance may affect hiring outcomes at least in part through the evaluations that intermediaries write about candidates.

There is ample evidence that physical attractiveness is rewarded by co-workers, supervisors, and customers \citep{Babin_et_al_2020,Stinebrickner_et_al_2019}.
Attractive individuals receive more favorable feedback and generate more positive interactions \citep{Jackson_et_al_1995,Langlois_et_al_2000}, and good looks can raise confidence and effectiveness in tasks that require human interaction
\citep{Mobius_Rosenblat_2006,Judge_et_al_2009}. Within academia, \citet{Fidrmuc_Paphawasit_2018} and \citet{Hale_et_al_2023} document associations between appearance, research productivity, and placement among economists.

To our knowledge, ours is the first study to document a direct association between attractiveness and reference letters. This is important because the literature suggests that attractiveness may operate as a signaling device. In other words, people use good looks as a proxy for competence.\footnote{For example, \citet{Deryugina_Shurchkov_2015} report that the laboratory beauty premium vanishes in tasks where productivity is directly measured and revealed to employers.} Our findings suggest that third-party evaluations do not fully eliminate this form of statistical discrimination.

Our paper also makes a methodological contribution to the measurement of attractiveness. While most of the literature has used human raters, there are a few prior attempts at using computational methods.\footnote{Researchers have of course been aware of the perils of using human judgment. Among the attempts to improve on this technique is the use of incentivized coordination games \citep[e.g.,][]{Babin_et_al_2020}, where raters are not asked to directly assess physical attractiveness but instead are asked to guess what they think other raters will think on average (i.e., a ``beauty contest'' design).} \citet{Peng_et_al_2020} extract multiple facial landmarks from face images, which they then use to train a random forest. While the training requires human-labeled examples, the approach is scalable, addressing one of the limitations of human ratings. Several recent papers use vision models to obtain embeddings from photos, which are then used to predict known characteristics of the individuals.\footnote{See, among others, \citet{Schroff_et_al_2015}, \citet{Karkkainen_Joo_2021}, \citet{Dosovitskiy_et_al_2021}, \citet{Kosinski_2021}, and \citet{Oquab_et_al_2023}.} Finally, \citet{Deng_Li_Zhou_2020} exploit the fact that celebrities are a distinct group of highly attractive people to generate fake photographs of job market applicants.\footnote{Specifically, they ``fuse'' real photos with the photo of a celebrity to enhance the applicant's attractiveness.} To our knowledge, ours is the first paper to use deep vision embeddings to measure physical attractiveness.

The paper proceeds as follows. Section~\ref{sec:methods} describes the method and formally defines the RVE score. Section~\ref{sec:validation} validates it against human ratings and examines the geometry of the embedding space. Section~\ref{sec:data} describes the academic job-market data.
Section~\ref{sec:results} presents the results on outcomes and on reference letters, and
Section~\ref{sec:gender} examines heterogeneity by gender. Section~\ref{sec:conclusion} concludes.

\section{A New Measure of Physical Attractiveness}\label{sec:methods}

This section provides a detailed account of the steps necessary to compute the RVE score. We provide a formal definition and explain the alternative possible specifications.

\subsection{From Photograph to Vector}\label{subsec_pipeline}

In this paper we introduce a new method for the measurement of physical attractiveness. In a nutshell, our approach to measure attractiveness involves calculating the proximity or similarity of a given subject's looks to people in a reference set that play the role of beauty ideal or benchmark. Therefore, our measure is essentially
relative (as in ``relative prices''), because a person's beauty score is only meaningful
relative to our given standard.

In order to measure proximity, we first use a deep vision model to obtain a dense numerical embedding representing the person's image. We similarly encode the images of the reference set using the same pipeline. Finally, we apply simple measures of vector distance to obtain our beauty score. We refer to this measure as the relative vision embedding (RVE) score.

\begin{figure}[hbt!]
  \centering
  \includegraphics[width=\textwidth]{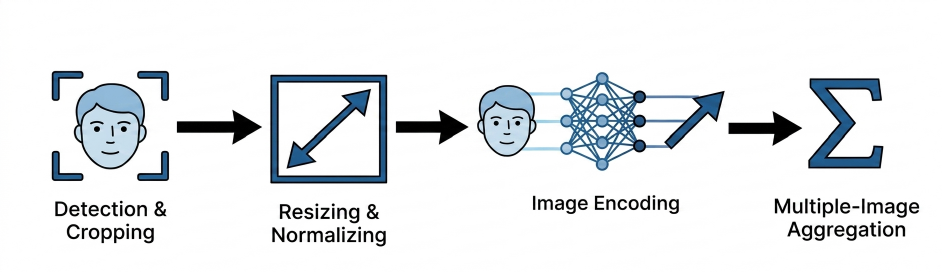}
  \caption{From Photograph to Embedding}\label{fig_pipeline}
  \vspace{-0.15in}
  {\footnotesize \flushleft Notes: 1. \texttt{RetinaFace} detects and crops the face; 2. the model-specific preprocessor resizes and normalizes the crop; 3. the vision model emits an embedding; 4. multiple photographs of the same individual are averaged. Image generated by Gemini AI, 3.1 Pro.}
\end{figure}

Figure~\ref{fig_pipeline} provides a schematic representation of the image processing and
encoding step. To make all images comparable, we focus on the person's face only.\footnote{Focusing on faces is the standard in the literature. Unfortunately, allowing for full-body images raises the issue of  the confounding effect of background and clothing, among other factors.} This is
done through a pre-processing step in which each image is submitted to \texttt{RetinaFace} \citep{Deng_et_al_2020}, a face-detection
and cropping model. Specifically, the detector returns bounding boxes and five-point landmarks for the largest face in each image.

Second, images are passed through a model-specific resizing and normalization
preprocessor. The details largely depend on the characteristics of the training data that was fed to each model. For example, the Inception model requires $299\times299$ pixel images and RGB values between 0 and 1.

Third, the image is submitted to the deep vision encoder, resulting in a single dense vector (embedding) per image. Finally, if multiple images are available for a single individual, the corresponding embeddings are averaged. By the end of this process, each individual is represented by a single vector $e_i$.

\subsection{Vision models}

We compute embeddings from four pre-trained models, listed in Table~\ref{tabl_models}, chosen to span different architectures, training objectives, and training corpora.

It is important to emphasize that we did not fine-tune any of the models. We consider this to be one major advantage of our approach. We take advantage of the model's pre-training without the need of any additional data gathering or further training.

From the models in the table, Inception v3 is the oldest, which explains why it is based on the (relatively old-fashioned) CNN architecture. It is also the smallest (by
number of parameters) by far. EVA and DINO are recently-released vision transformers. Their architecture is fundamentally the same as the now famous text transformers like BERT or GPT. These models first break an image into a series of patches, which are then tokenized and fed to the model as a sequence. Finally, Qwen3 is a large multimodal model with over 2B parameters. We only use the latter model's ability to embed images.\footnote{Multimodality---the ability to process both text and image inputs---is a common characteristic of state-of-the-art transformers.}

\begin{table}[htb!]
  \centering
  \caption{Pre-trained Deep Vision Models}\label{tabl_models}
  {\scriptsize
    \include{tables/tabl_models}
  }
\end{table}

\subsection{The RVE Score}

\begin{figure}[hbt!]
  \centering
  \includegraphics[width=0.8\textwidth]{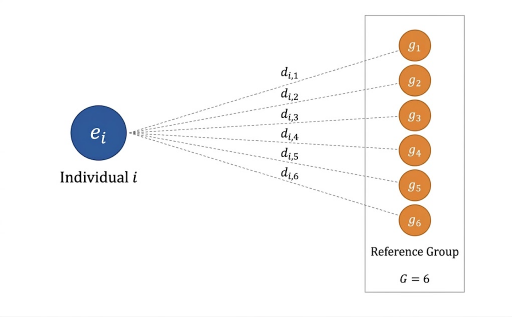}
  \caption{RVE Score Calculation}\label{fig_rve_calculation}
  \vspace{-0.15in}
  {\footnotesize \flushleft Notes: The RVE score is the average proximity from an individual's embedding to the embeddings of the reference set.}
\end{figure}

Once all images have been vectorized, calculating a beauty score is relatively
straightforward (see figure~\ref{fig_rve_calculation}). The RVE score is just the average proximity of an individual's embedding to the embeddings of the reference set. Formally, let the embeddings for the members of the reference set be denoted as $\mathbb{G}=\{g_1, g_2, \ldots, g_G\}$. For a given proximity metric $d(\cdot,\cdot)$ in the embedding space, the individual's RVE score is thus defined as

\begin{align}\label{eq_rve}
  b_i &= \frac{1}{G}\sum_{j=1}^{G} d(e_i,g_j)\\
\end{align}

We use a set of celebrities as the baseline reference set (see details below). For robustness, we consider three variants (subsets) of this approach: all-celebrities ($\mathbb{G}_{all}$), same-gender-celebrities ($\mathbb{G}_{gender}$), and same-race-and-gender-celebrities ($\mathbb{G}_{race\&gender}$).

\begin{figure}[hbt!]
  \centering
  \includegraphics[width=0.7\textwidth]{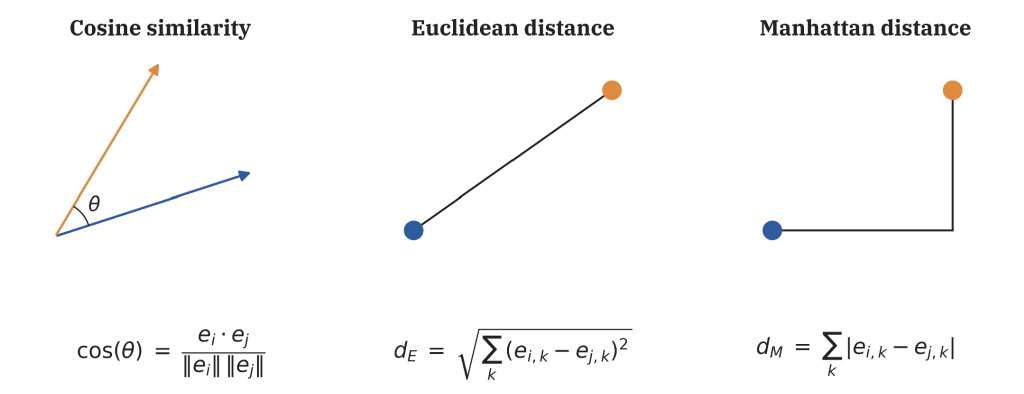}
  \caption{Proximity Metrics}\label{fig_distance_metrics}
  \vspace{-0.15in}
  {\footnotesize \flushleft Notes: Cosine similarity captures the angle between embedding vectors and is invariant to vector magnitude. Euclidean distance is the ``straight line'' distance. Manhattan distance sums absolute coordinate differences.}
\end{figure}

We also report results for three alternative  proximity metrics: cosine similarity, Euclidean distance, and Manhattan distance (see figure~\ref{fig_distance_metrics}). Because the Euclidean and Manhattan distance measure dissimilarity, we take the
sign-reversed distance and normalize the measures to be in the $[0,1]$ interval.

\subsection{Celebrity Data (Reference Set)}\label{subsec_celebrities}

The reference set plays a central role in our method. It is populated with photographs of celebrities collected from Ranker,\footnote{\url{www.ranker.com}, accessed February 7th 2026.} a public website on which categories of people are ordered by public user voting. We downloaded the attractiveness rankings split by race and gender, using the top 100 entries of each list.\footnote{The exception is Hispanic females, whose list only had 76 elements. The lists for Whites were created by removing non-Whites from a general ranking. Overall, we have 776 celebrities in 8 gender-race categories (Male/Female and White/Black/Hispanic/Asian).} The RVE
score does not use a celebrity's rank position or vote counts, and each face enters the reference set with equal weight. Only
membership in the top 100 matters. We collected 2 photographs per celebrity from Ranker and applied the pipeline described in
Section~\ref{subsec_pipeline}.

The reference set has an objective claim to represent the very top of the
attractiveness distribution, because its members are doubly selected. Physical
attractiveness is highly rewarded in entertainment and media, so attractive individuals are more likely to enter public prominence in the first place. Ranker's
attractiveness-specific rankings further select within that already-selected pool.
Each step conditions on looks, and their composition places the benchmark in the
high-attractiveness region of face space.

The benchmark is also reasonably large, at almost 800 faces, so individual faces carry little weight. The RVE score is a mean over the full set, and no single celebrity drives
it. This makes the measure easy to stress-test. We compute scores against the
pooled set ($\mathbb{G}_{all}$), same-gender celebrities ($\mathbb{G}_{gender}$),
and same-race-and-gender celebrities ($\mathbb{G}_{race\&gender}$), and we
re-estimate rankings on random subsets of each. Because the benchmark is fixed and
public, the construction is fully reproducible, and anyone with the same
embeddings obtains identical scores for identical images.

Celebrity photographs are taken under favorable conditions and professionally
styled, so the benchmark carries a systematic photo-quality advantage over the
faces being scored. Because this advantage is a fixed feature of the benchmark, it
shifts the RVE score of every subject in the same direction. In a cross-sectional
comparison it is absorbed by the intercept and does not affect the ranking of
subjects, which is what the estimates depend on.

Finally, collecting the reference set is relatively costless and can be scaled up
if needed. The lists are already split by race and gender, which simplifies the
construction of group-specific scores.

\section{Validation}\label{sec:validation}

In this section we document two facts about the RVE score. First, it is closely
aligned with average human attractiveness ratings. Second, a visualization of the
embedding space shows that the embeddings capture information relevant to physical
attractiveness. Specifically, celebrities form a distinct cluster in the embedding
space, and the measure assigns them high attractiveness scores when each celebrity
is scored against a reference set that excludes them.

\subsection{The RVE Score vs.\ Human Rated Attractiveness}\label{subsubsec_hrs}

We test whether RVE scores are comparable to scores produced by humans. For this
purpose, we use the Chicago Face Database 3.0 (CFD): a publicly available dataset
with a large number of images rated on attractiveness by volunteers.

The CFD is particularly well suited to this exercise because its photographs are
highly standardized: all subjects are photographed against a white background,
centered and facing the camera, wearing a plain white t-shirt, and without
special styling or makeup. All subjects consented to the public release of their
images. These features remove the photographic and stylistic variation that
would otherwise confound a comparison between model-based and human-based
attractiveness scores.

We restrict our exercise to the main CFD sample, consisting of 597 individuals
for whom self-reported gender and race identifiers are available. For each
individual we have one standardized (profile-picture-like) neutral-expression
photograph. In addition, each individual was rated on attractiveness (on a 1--7
scale) by 28 human raters. We use the mean rating, which we refer to as the
human-rated score (HRS). Appendix~\ref{app_cfd} shows the top- and bottom-rated
individuals for each gender--race group in the sample.

Table~\ref{tabl_rve_human} reports the Spearman (rank) and Pearson correlations
for all the RVE variants against the HRS. In general, EVA-02 is the model that
performs closest to the human raters. It attains the highest correlation
($\rho=0.484$) using the same-gender celebrity reference group and Manhattan
distance. In contrast, Qwen3 performs poorly and leads to substantively different rankings relative to the humans.

\begin{table}[htb!]
  \centering
  \caption{RVE Scores and Human Attractiveness Ratings}\label{tabl_rve_human}
  {\tiny
    \include{tables/tabl_rve_human}
  }
\end{table}

One option would be to select the RVE variant whose ratings are most closely
aligned with humans. This approach, however, risks overfitting to the CFD and
may not generalize to other evaluation settings. Instead, we adopt a compromise:
we drop Qwen3 and construct an ensemble from the embeddings of the remaining
three models.\footnote{We believe the under-performance of Qwen3 is due to its multimodal architecture. Further investigating this hypothesis is left for future research.}

\begin{table}[htb!]
  \centering
  \caption{A 3-Model Ensemble}\label{tabl_ensemble}
  {\tiny
    \include{tables/tabl_ensemble}
  }
\end{table}

Table~\ref{tabl_ensemble} (panel A) presents similarly obtained correlations for
the resulting three-model ensemble. The ensemble performs close to EVA-02. We
expect that, in addition, the ensemble will be more robust to images that differ
somewhat from the CFD.

Panel B of Table~\ref{tabl_ensemble} presents correlations for the different
demographic sub-groups of the CFD. The within-group correlations are markedly
higher than the overall correlation. The reason is that the positive
within-group relationship is offset by between-group differences in means:
women receive higher HRS but lower RVE scores, so the pooled correlation is
attenuated.\footnote{This pattern is often referred to as Simpson's paradox.} Table~\ref{tabl_cfd_means}
presents the corresponding sample means.

\begin{table}[htb!]
  \centering
  \caption{CFD Sample Means}\label{tabl_cfd_means}
  {\tiny
    \include{tables/tabl_cfd_means_v1}
  }
\end{table}

The main takeaway from the table is that human raters tend to give higher scores to women, whereas RVE scores for men are somewhat higher than for women. One possible interpretation is a rater-side positivity bias toward women---social desirability or chivalry effects---which affects human raters but not the vision models. In this respect the RVE score has a distinct advantage over human ratings: it is applied mechanically and identically to every image, free of rater-side scale-use and social-desirability effects.

\begin{figure}[htbp]
  \centering
  \includegraphics[width=0.6\textwidth]{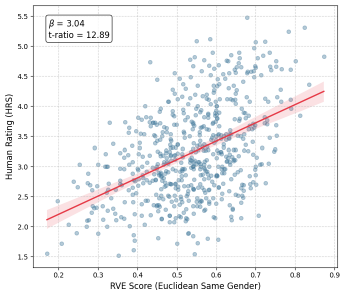}
  \caption{Human-rated attractiveness (HRS) against the ensemble RVE score}
  \label{fig_hrs_rve}
\end{figure}

\begin{table}
  \centering
  \caption{HRS Regression}\label{tabl_hrs_rve_reg}
     {\small
      \include{tables/cfd_regressions/tab_hrs_rve_regression}

     }
\end{table}

Figure~\ref{fig_hrs_rve} presents the joint distribution of the ensemble-based RVE score and the HRS for the whole CFD sample, together with a linear fit. The relationship is clearly positive. Table~\ref{tabl_hrs_rve_reg} presents regression estimates. The unconditional relationship is highly statistically significant, and it remains so---with a larger coefficient---after controlling for gender and race. The increase in the coefficient upon adding controls reflects the suppression pattern documented in Table~\ref{tabl_cfd_means}: because women receive higher HRS but lower RVE scores, the demographic controls absorb the between-group component and leave a
sharper within-group relationship.

\subsection{Visualisation of the Embedding Space}\label{subsubsec_visualization}

\begin{figure}[htbp]
  \centering
  \includegraphics[width=0.85\textwidth]{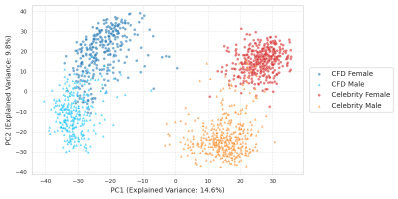}
  \caption{Embedding Space Visualization (PCA)}
  \label{fig_pca}
\end{figure}

In order to gain insight into the workings of the deep vision models, we present a two-dimensional visualization of the workings of the RVE measure. For this purpose, Figure~\ref{fig_pca} displays the first two principal components of the vector embeddings. The figure clearly shows that 1) the embeddings of celebrities and CFD subjects are separable; and 2) males and females are also quite distinct.

\begin{figure}[htbp]
  \centering
  \includegraphics[width=0.7\textwidth]{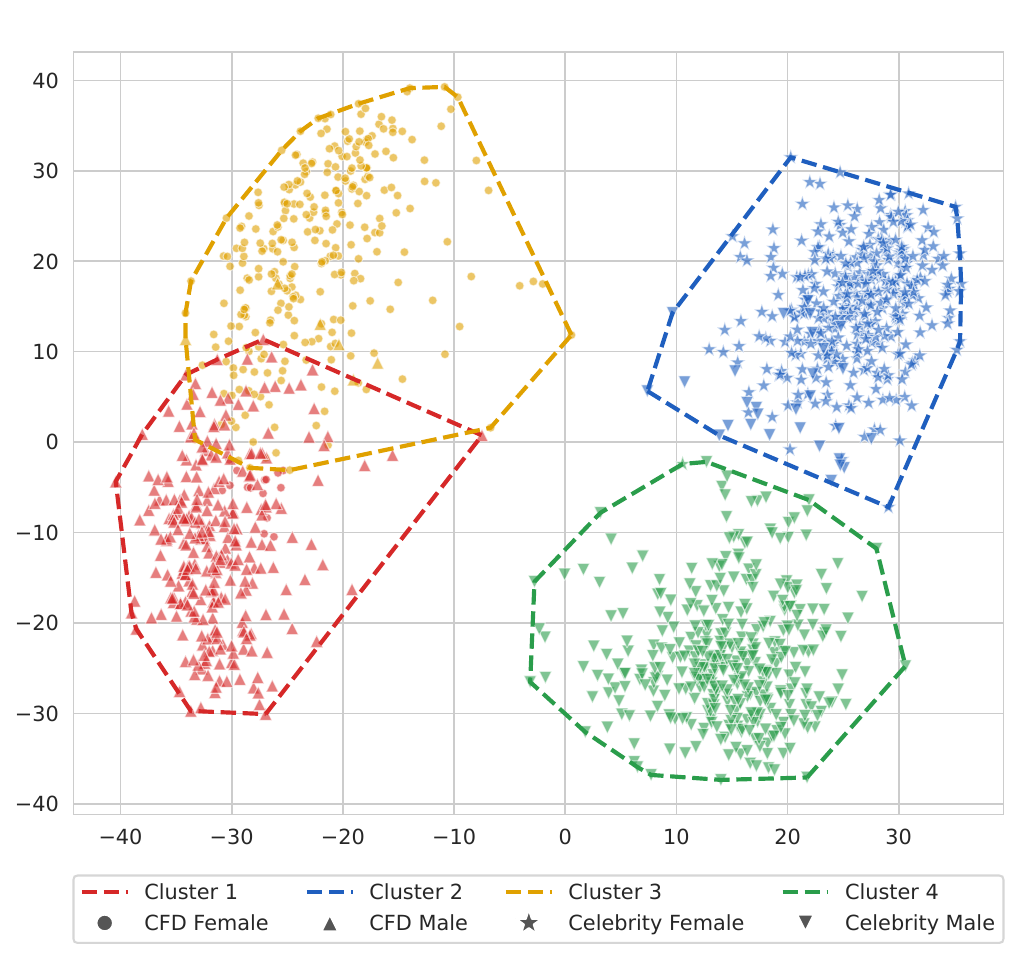}
  \caption{$K$-means Clustering of Ensemble Embeddings}
  \label{fig_kmeans}
\end{figure}

To reinforce this point, we applied $K$-Means clustering to the concatenated embeddings. Specifically, we let the algorithm search for four clusters. Figure~\ref{fig_kmeans} overlays the PCA projections with the resulting clusters. Celebrities and CFD subjects fall into different clusters entirely. The clusters also separate males and females well: Cluster 1 is dominated by male CFD subjects (94.7\%), Cluster 2 by female celebrities (92.8\%), Cluster 3 by female CFD subjects (98\%), and Cluster 4 consists almost purely of male celebrities (99.7\%). The embedding therefore separates the populations and the demographic groups that enter the reference sets.

\begin{figure}[htbp]
  \centering
  \includegraphics[width=0.7\textwidth]{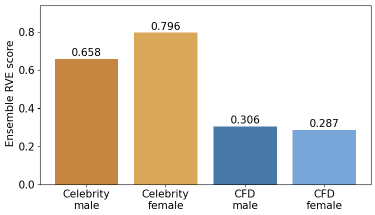}
  \caption{Average RVE scores: CFD versus celebrity reference set.}
  \label{fig_rve_groups}
\end{figure}

To compute RVE scores for celebrities, we exclude each celebrity from their own reference set and score them against the remaining celebrities. Celebrities score much higher than CFD subjects (Figure~\ref{fig_rve_groups}), which confirms that the reference set is internally coherent and that the measure ranks the reference population at the top of the scale. Because the reference set consists of celebrities, this comparison is a consistency check on the construction rather than an out-of-sample test.

\section{The Academic Job Market}\label{sec:data}

In this section, we use the RVE score to investigate the role of physical attractiveness in the academic job market for economists. We examine two related questions. First, do RVE scores predict job-market placement? Second, do
RVE scores predict the content of the reference letters that candidates receive---their length and their sentiment? We show that the evidence points to letters as a channel through which
appearance reaches the hiring decision.

\subsection{Setting and Data}

Our starting point is the hand-collected list of job market candidates compiled by
\citet{Slonimczyk_2025}. The dataset includes 645 fresh PhDs who applied for a
tenure-track position at a research-oriented economics department in Russia during the period 2013--2021. The department recruits internationally, so the applicant pool spans
all the demographic groups in our data. The data includes information on the candidate's PhD institution, their publication
record, and the length of their studies. Finally, we have detailed information on the length and sentiment of the reference letters submitted on behalf of the candidates, as well as information on the letter writers.\footnote{We use the pre-trained language model prompt-completion approach to extracting sentiment from the reference letters \citep[see][]{Slonimczyk_2025}.}

We supplement these data with hand-collected photos for almost all of the job market candidates. The images come from open sources like LinkedIn profiles, personal websites, and university staff profiles, all of which are routinely consulted by hiring committees. Committees also meet short-listed candidates in
person before making an offer. Overall, we were able to collect at least one photo for 630 of the job market candidates. Many candidates had multiple usable images, so we included all of them in our dataset and averaged their embeddings at the candidate level. In summary, we collected 1,394 photos, an average of 2.21 per candidate.

The sample is quite unbalanced in terms of gender and race. There are 465 males (73.8
percent) and 165 females (26.2 percent). Over 60 percent of the sample is White. Moreover,
the very small number of Black candidates makes inference for that group impossible, but we
still report their results for the sake of completeness. Table~\ref{tabl_sample_composition} lays out the detailed demographic composition of the sample.

\begin{table}[htb!]
  \centering
  \caption{Sample Composition}\label{tabl_sample_composition}
  {\tiny
    \include{tables/tabl_sample_composition}
  }
\end{table}

\subsection{Descriptive Statistics}\label{subsec_descriptives}

We present statistics for RVE scores obtained using the 3-model ensemble, Euclidean distance, and same-gender celebrities as the reference group. Results
for other specifications are similar and available in the online appendix.

\begin{table}[htb!]
  \centering
  \caption{RVE Ensemble Scores}\label{tabl_rve_jm}
  {\small
    \include{tables/tabl_rve_jm}
  }
\end{table}

\begin{figure}[hbt!]
  \centering
  \includegraphics[width=0.7\textwidth]{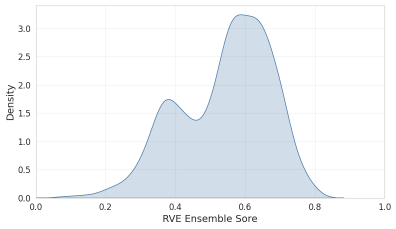}
   \caption{RVE Scores Kernel Density Estimate (all candidates)}\label{fig_rve_kde}
\end{figure}

Table~\ref{tabl_rve_jm} shows the distribution of RVE scores for the sample of job market
candidates. While there are some small differences among demographic groups, they are
overshadowed by the remarkable spread in the distributions within groups.
Figure~\ref{fig_rve_kde} shows a kernel estimate of the RVE score distribution. The
distribution is bimodal, suggestive of a split into below-average and above-average looks.

\subsection{Results}\label{sec:results}

In this section, we examine the relationship between attractiveness---as measured by
the RVE score---and both job-market employment outcomes and reference-letter contents. We
consider outcomes first and then turn to letters.

\subsubsection{Job-market Outcomes}\label{subsec_outcomes}

Following the extensive literature showing that ``beauty matters'', we explore whether
our RVE score is predictive of job market outcomes. Following our earlier work \citep{Slonimczyk_2025}, we define ``success'' in the academic job market with a binary indicator of whether the candidate is hired by either a top-100 research academic institution or a top-50 policy-oriented institution.\footnote{Academic institutions are classified using the Tilburg ranking. For policy institutions, we use the RePEC ranking.} We consider job placement by September following the corresponding job market season. We also explore four alternative outcomes for robustness. Table~\ref{tabl_outcome_defs} has the detailed definitions.

\begin{table}[htb!]
  \centering
  \caption{Outcome Definitions}\label{tabl_outcome_defs}
  {\small
    \input{tables/tabl_outcome_defs}
  }
\end{table}

\begin{figure}[hbt!]
  \centering
  \includegraphics[width=0.97\textwidth]{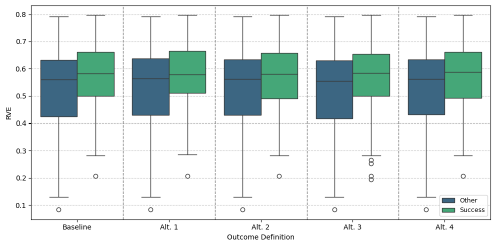}
  \caption{Average RVE Scores by Job Market Outcomes}\label{fig_rve_outcomes}
\end{figure}

Figure~\ref{fig_rve_outcomes} shows the distribution of RVE scores by job market outcome.
Regardless of the precise definition, successful candidates score higher on our measure of attractiveness. To explore whether these differences are statistically significant, we
estimate the linear probability model:

\begin{align}\label{eq_success}
  P(\mathrm{success} \mid RVE, X) &= \beta_0 + \beta_1 RVE + X^\prime\theta + \epsilon
\end{align}

where the $X$ vector contains a set of control variables. We report HC3 robust standard errors.

\begin{table}[htbp]
  \centering
  \caption{Job-market Outcome and Attractiveness}
  \label{tabl_LM1_RVE_reg}
  \include{tables/tabl_LM_outcome1_RVE}
\end{table}

\begin{table}[htbp]
  \centering
  \caption{Alternative Outcome Definitions and Attractiveness}
  \label{tabl_LM1-5_RVE_reg}
  \include{tables/tabl_LM_outcome1-5_RVE}
\end{table}

Table~\ref{tabl_LM1_RVE_reg} shows regression results. Column 1 verifies that the differences
in means in figure~\ref{fig_rve_outcomes} are indeed statistically significant. In column 2,
we add a number of relevant control variables describing the candidate's characteristics
(gender, race, publications). In general, the relationship between RVE and outcomes remains
significant. Column 3 further adds characteristics of the PhD program (whether the program is U.S.-based and its ranking tier) and the job-market year. The coefficients are still positive, but smaller and marginally significant. Finally, columns 4--6 add reference letter characteristics to the equation. Consistent with our earlier findings \citep{Slonimczyk_2025}, average letter length (word count) and sentiment are both positively associated with the probability of success. In contrast, the RVE score coefficient is now small and no longer statistically significant.

Table~\ref{tabl_LM1-5_RVE_reg} shows regression results for the full specification, comparing the baseline to other outcome definitions. The results are fundamentally identical. Once reference letter characteristics are included, the RVE score does not predict job-market success.

Our reading of these results is deliberately restrained. The sign is consistently positive,
which is consistent with a beauty premium in academic hiring. But the association is not
robust to every specification and does not survive at conventional significance levels once the full control set is included in some columns. We therefore describe Finding~1 as a
\emph{positive but not fully robust} association. There are three reasons not to over-read the imprecision. The outcome is coarse; the sample is small by the standards of the beauty literature; and appearance is plausibly a second-order determinant of academic hiring relative to research output, field, and Ph.D. pedigree.

A more important reason not to read the regression results as evidence that appearance does not matter is that reference letters are a \emph{bad control} for this question: they are themselves potentially an outcome of attractiveness. Comparisons of individuals with different RVE scores but the same letter characteristics are not apples-to-apples comparisons.

\subsubsection{Reference Letters}\label{subsec_letters}

We move to consider the relationship between recommendation letters and attractiveness. To our knowledge, this is the first study to link a measure of physical attractiveness to the content of recommendation letters.

We conduct the analysis at the letter level and cluster standard errors at the candidate level, since multiple letters describe the same candidate and share unobserved candidate characteristics.

\begin{figure}[htbp]
  \centering
  \includegraphics[width=0.9\textwidth]{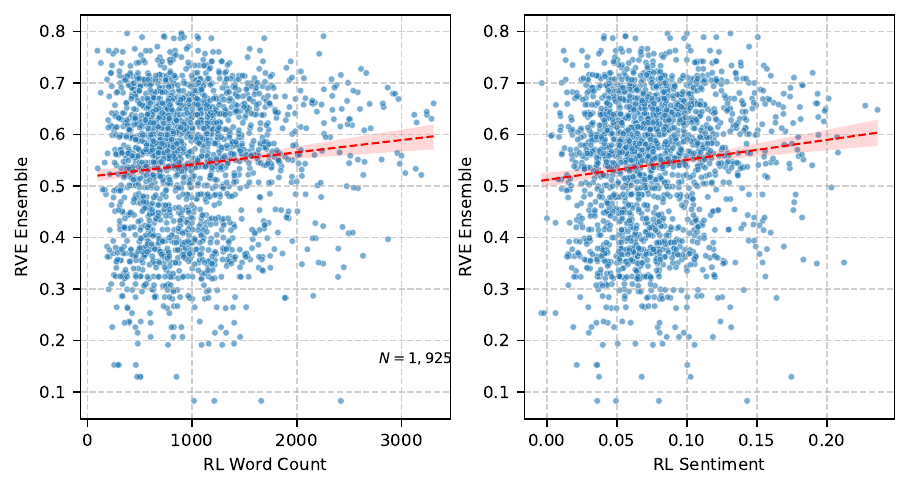}
  \caption{RVE score against reference-letter word count and sentiment.}
  \label{fig:rl-scatter}
\end{figure}

Figure~\ref{fig:rl-scatter} plots RVE scores against letter word count and letter sentiment. In both cases, there is a clear positive relationship.

Table~\ref{tab:writers} reports the characteristics of the letter writers that we include as controls to ensure that the estimated relationship is not driven by the propensity of more distinguished letter writers to write about better-looking candidates.

\begin{table}[htbp]
  \centering
  \caption{Characteristics of letter writers.}
  \label{tab:writers}
  \include{tables/tabl_writers_desc_stats_presentation}
\end{table}

Table~\ref{tab:rl-reg} reports the regressions. We estimate

\begin{equation}
Y_{i\ell} \;=\; \alpha + \beta\,\RVE_i + X_i'\gamma + W_{i\ell}'\delta
+ \varepsilon_{i\ell},
\end{equation}

where $Y_{i\ell}$ is a characteristic of the letter written by $\ell$ about $i$---word
count and sentiment score---$X_i$ is a vector of candidate controls, and $W_{i\ell}$ contains letter-writer characteristics. The estimated
$\beta$ is positive and significant for word count and for sentiment. These estimates are robust to the full control set.

\begin{table}[htbp]
  \centering
  \caption{Regressions: reference-letter characteristics and attractiveness.}
  \label{tab:rl-reg}
  \include{tables/tabl_reg_RL_rve_presentation}
\end{table}

The economic content is straightforward. A more attractive candidate receives a longer
letter and a letter whose language is more favorable, holding constant the characteristics of both the candidate and the writer. If search committees read letters as informative
signals about fit and promise, then letters are a channel through which appearance reaches
the hiring decision.

\begin{figure}[htbp]
  \centering
  \includegraphics[width=0.5\textwidth]{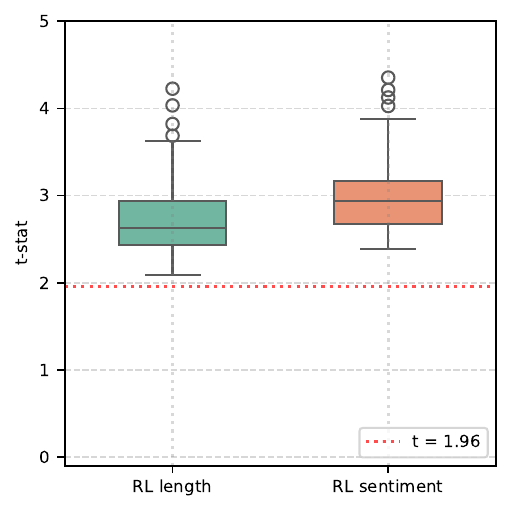}
  \caption{Distribution of $t$-statistics on the RVE coefficient across 45 alternative
  measures and specifications.}
  \label{fig:tstats}
\end{figure}

These results are stable across specifications. Figure~\ref{fig:tstats} reports the distribution of $t$-statistics on the RVE coefficient across 45 alternative specifications: different distance metrics, different reference-set definitions, and different regression controls. In 100\% of specifications the coefficient is significant at least at the 5\% level.

\subsubsection{Heterogeneity by Gender}\label{sec:gender}

\begin{figure}[htbp]
  \centering
  \includegraphics[width=0.9\textwidth]{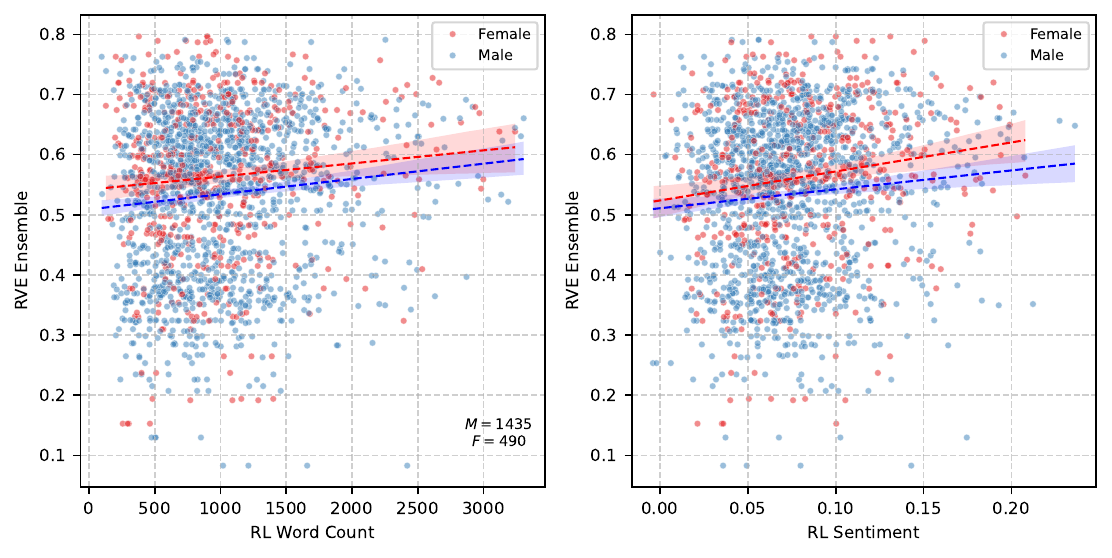}
  \caption{Gender, Reference-Letter Characteristics and RVE Score}
  \label{fig:rl-gender}
\end{figure}

The beauty premium is frequently found to differ by gender, and there is a well-documented tendency for evaluations of women to weight appearance more heavily than evaluations of men. We therefore estimate the reference-letter specifications including interactions by gender.
Figure~\ref{fig:rl-gender} plots RVE against letter characteristics, and
Table~\ref{tab:rl-gender} reports the regressions with interactions.

\begin{table}[htbp]
  \centering
  \caption{Gender Reference-Letters and RVE Score: Gender Heterogeneity}
  \label{tab:rl-gender}
  {\tiny
  \include{tables/tabl_reg_RL_rve_gender_presentation}

  }
\end{table}

\begin{figure}[htbp]
  \centering
  \includegraphics[width=0.5\textwidth]{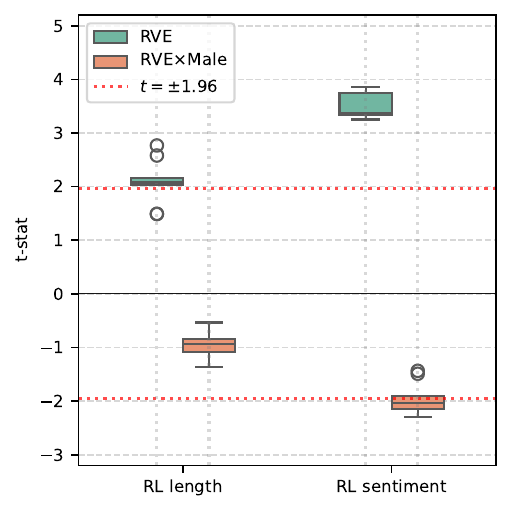}
  \caption{Distribution of $t$-statistics across 9 alternative specifications, with gender interactions.}
  \label{fig:tstats-gender}
\end{figure}

The sentiment gradient is steeper for female candidates: the coefficient on the RVE score in
the sentiment equation is 0.044 for women and 0.011 for men, a difference
that is significant at conventional levels. The word-count gradient
is not significantly different across genders. Figure~\ref{fig:tstats-gender}
reports the corresponding distribution of $t$-statistics across 9 alternative
specifications. The gender difference in the
sentiment gradient is not an artifact of a single model choice but is also not always statistically significant.

A gender difference in the slope of the relationship admits several interpretations: it may reflect differential weighting of appearance in judgments about women, or it may reflect gender differences in how attractiveness correlates with other traits that letter writers respond to, such as confidence or social skills. What the result does establish is that the appearance gradient in letters is not a uniform phenomenon.

\section{Conclusion}\label{sec:conclusion}

We propose the Relative Vision Embedding score as a measurement instrument for physical
attractiveness. The score is the average proximity (e.g., the sign-reversed distance) from a person's face
embedding to a reference group of celebrities, computed in the embedding space of a pre-trained deep vision model. It requires no training data, no hand-crafted features, and no human raters; it is continuous, reproducible, and cheap enough that an entire robustness grid costs minutes. It correlates strongly with human attractiveness ratings in the Chicago Face Database, especially within demographic groups, and rankings of faces are similar across architectures.

Applied to 630 fresh Ph.D.s in economics, the score is positively associated with job-market success, though not decisively so, and strongly and robustly associated with the content of reference letters: better-looking candidates receive longer and more favorable letters, with a steeper sentiment gradient for women. The pattern is consistent with appearance shaping academic hiring in part through the letters candidates receive.

Two directions for further research seem most promising. First, because the measure is portable, it can be applied to archival photographs, to other national labor markets, and to other hiring settings at essentially no cost, which makes cross-study comparability possible for the first time. Second, the letter-content results suggest that the beauty premium in academia may operate
less through the interview and more through the file, which is a testable proposition with
better data on letter content.

\clearpage

\singlespacing
\bibliographystyle{kbib}
\bibliography{bibdata}
\addcontentsline{toc}{section}{References}
\clearpage

\begin{center}
  {\Large\bfseries Appendix}
\end{center}
\appendix
\setcounter{section}{0}

\section{Celebrities}\label{app_celebrities}

The celebrity panel was extracted from \url{www.ranker.com}. Specifically, we relied on the
following user-ranked lists:

\bigskip
\noindent\textbf{Asian Males}\\
\url{https://www.ranker.com/list/hottest-asian-men/calistylie}
\begin{center}
  \includegraphics[width=0.95\textwidth]{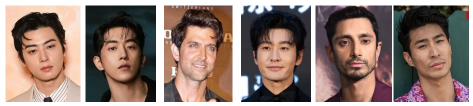}
\end{center}

\noindent\textbf{Asian Females}\\
\url{https://www.ranker.com/list/most-beautiful-asian-women/calistylie}
\begin{center}
  \includegraphics[width=0.95\textwidth]{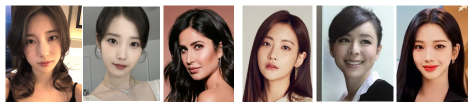}
\end{center}

\noindent\textbf{Black Males}\\
\url{https://www.ranker.com/list/hottest-black-men/calistylie}
\begin{center}
  \includegraphics[width=0.95\textwidth]{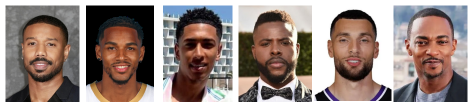}
\end{center}

\noindent\textbf{Black Females}\\
\url{https://www.ranker.com/list/beautiful-black-celebrity-women/greg}
\begin{center}
  \includegraphics[width=0.95\textwidth]{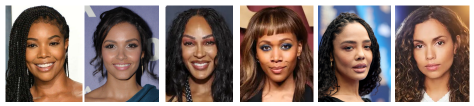}
\end{center}

\noindent\textbf{Latin Males}\\
\url{https://www.ranker.com/list/hottest-latin-men/calistylie}
\begin{center}
  \includegraphics[width=0.95\textwidth]{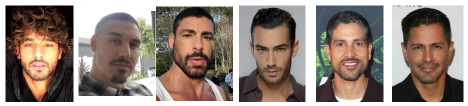}
\end{center}

\noindent\textbf{Latin Females}\\
\url{https://www.ranker.com/list/most-beautiful-latina-women/calistylie}
\begin{center}
  \includegraphics[width=0.95\textwidth]{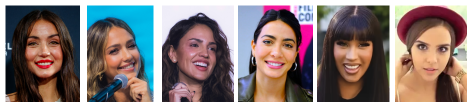}
\end{center}

\noindent\textbf{White Males}\\
\url{https://www.ranker.com/crowdranked-list/hottest-male-celebrities}
\begin{center}
  \includegraphics[width=0.95\textwidth]{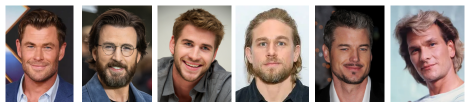}
\end{center}

\noindent\textbf{White Females}\\
\url{https://www.ranker.com/crowdranked-list/most-beautiful-women-in-hollywood}
\begin{center}
  \includegraphics[width=0.95\textwidth]{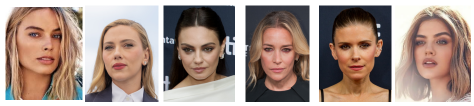}
\end{center}

\clearpage
\section{The Chicago Face Dataset (CFD)}\label{app_cfd}

Top- and bottom-rated photographs according to their average human-rated score (HRS):

\subsection{Top-ten-ranked subjects according to human raters.}

\begin{center}
  \includegraphics[width=\textwidth]{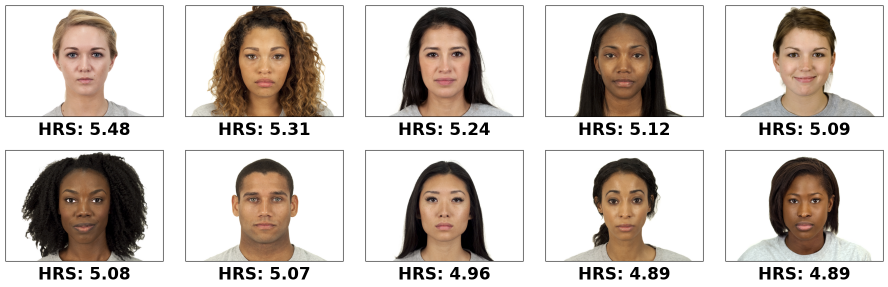}
\end{center}

\subsection{Bottom-ten-ranked subjects according to human raters.}

\begin{center}
  \includegraphics[width=\textwidth]{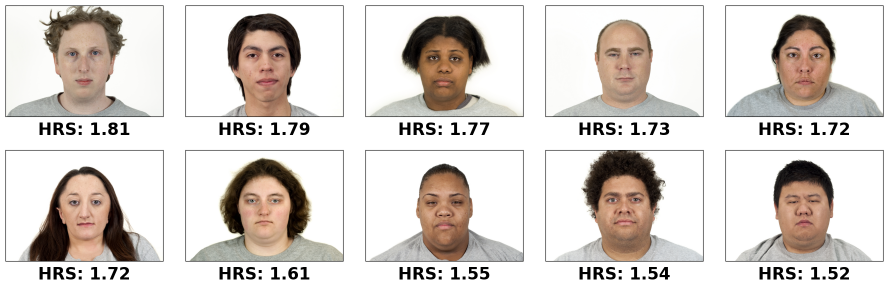}
\end{center}

\end{document}

%% file: tables/tabl_models.tex
\setlength{\tabcolsep}{2pt}
\begin{tabular}{llccll}
\toprule
Model & Architecture & \begin{tabular}[c]{@{}c@{}}Embedding\\ dimension\end{tabular} & Parameters & Source & Developer \\
\midrule
DINOv3 ViT-L/16        & Vision Transformer (ViT)  & 1,024 & 300M & HuggingFace & Meta AI (2025) \\
EVA-02 ViT-L           & ViT                       & 1,024 & 304M & HuggingFace & BAAI (2023)    \\
Qwen3-VL-Embedding-2B  & Multimodal Transformer    & 2,048 & 2B   & HuggingFace & Alibaba (2025) \\
Inception v3           & Convolutional Neural Network (CNN) & 2,048 & 24M & Tf Hub & Google (2016) \\
\bottomrule
\end{tabular}

%% file: tables/tabl_rve_human.tex
\setlength{\tabcolsep}{3.5pt}
\begin{tabular}{llccccccccc}
\toprule
 & & \multicolumn{3}{c}{Same Race \& Gender} & \multicolumn{3}{c}{Same Gender} & \multicolumn{3}{c}{All Celebrities} \\
\cmidrule(lr){3-5}\cmidrule(lr){6-8}\cmidrule(lr){9-11}
Model & & Cosine & Euclid. & Manh. & Cosine & Euclid. & Manh. & Cosine & Euclid. & Manh. \\
\midrule
DINOv3    & $\rho$ & 0.140 & 0.168 & 0.153 & 0.135 & 0.146 & 0.125 & \textbf{0.359} & 0.203 & 0.178 \\
          & $r$    & 0.147 & 0.177 & 0.160 & 0.137 & 0.166 & 0.145 & \textbf{0.385} & 0.227 & 0.193 \\
\addlinespace
Inception & $\rho$ & 0.157 & 0.289 & 0.272 & 0.217 & \textbf{0.334} & 0.321 & 0.003 & 0.129 & 0.133 \\
          & $r$    & 0.200 & 0.295 & 0.281 & 0.244 & \textbf{0.330} & 0.317 & $-$0.015 & 0.124 & 0.120 \\
\addlinespace
EVA-02    & $\rho$ & 0.328 & 0.373 & 0.385 & 0.422 & 0.469 & \textbf{0.484} & 0.360 & 0.365 & 0.389 \\
          & $r$    & 0.355 & 0.398 & 0.410 & 0.459 & 0.507 & \textbf{0.519} & 0.379 & 0.387 & 0.405 \\
\addlinespace
Qwen3-VL  & $\rho$ & 0.050 & 0.081 & 0.068 & 0.036 & 0.064 & 0.053 & \textbf{0.096} & 0.090 & 0.077 \\
          & $r$    & 0.066 & 0.096 & 0.081 & 0.054 & 0.083 & 0.070 & \textbf{0.106} & 0.098 & 0.090 \\
\bottomrule
\multicolumn{11}{p{2.6in}}{Notes: in bold are the highest row-wise correlations.}\\
\end{tabular}

%% file: tables/tabl_ensemble.tex
\setlength{\tabcolsep}{3.5pt}
\begin{tabular}{lrlccccccccc}
\toprule
 & & & \multicolumn{3}{c}{Same Race \& Gender} & \multicolumn{3}{c}{Same Gender} & \multicolumn{3}{c}{All Celebrities} \\
\cmidrule(lr){4-6}\cmidrule(lr){7-9}\cmidrule(lr){10-12}
Subgroup & $n$ & & Cosine & Euclid. & Manh. & Cosine & Euclid. & Manh. & Cosine & Euclid. & Manh. \\
\midrule
\multicolumn{12}{l}{\textit{Panel A. All CFD individuals}} \\
\addlinespace
All & 597 & $\rho$ & 0.29 & 0.40 & 0.40 & 0.28 & 0.42 & 0.42 & 0.38 & 0.37 & 0.37 \\
    &     & $r$    & 0.30 & 0.41 & 0.41 & 0.33 & 0.47 & 0.46 & 0.39 & 0.39 & 0.39 \\
\midrule
\multicolumn{12}{l}{\textit{Panel B. CFD subgroups}} \\
\addlinespace
All females & 307 & $\rho$ & 0.63 & 0.65 & 0.65 & 0.73 & 0.72 & 0.73 & 0.51 & 0.54 & 0.54 \\
            &     & $r$    & 0.64 & 0.66 & 0.66 & 0.73 & 0.73 & 0.73 & 0.52 & 0.55 & 0.55 \\
\addlinespace
\ \ Female $\times$ Asian    &  57 & $\rho$ & 0.66 & 0.62 & 0.66 & 0.70 & 0.67 & 0.70 & 0.51 & 0.55 & 0.57 \\
                             &     & $r$    & 0.66 & 0.63 & 0.66 & 0.69 & 0.67 & 0.71 & 0.50 & 0.54 & 0.56 \\
\addlinespace
\ \ Female $\times$ African  & 104 & $\rho$ & 0.74 & 0.73 & 0.73 & 0.75 & 0.74 & 0.73 & 0.54 & 0.56 & 0.53 \\
                             &     & $r$    & 0.75 & 0.74 & 0.75 & 0.75 & 0.75 & 0.75 & 0.52 & 0.57 & 0.54 \\
\addlinespace
\ \ Female $\times$ Hispanic &  56 & $\rho$ & 0.59 & 0.64 & 0.66 & 0.59 & 0.62 & 0.64 & 0.50 & 0.54 & 0.50 \\
                             &     & $r$    & 0.69 & 0.71 & 0.73 & 0.69 & 0.70 & 0.73 & 0.54 & 0.57 & 0.55 \\
\addlinespace
\ \ Female $\times$ White    &  90 & $\rho$ & 0.71 & 0.73 & 0.73 & 0.76 & 0.77 & 0.77 & 0.66 & 0.69 & 0.68 \\
                             &     & $r$    & 0.70 & 0.72 & 0.72 & 0.75 & 0.76 & 0.76 & 0.65 & 0.69 & 0.68 \\
\addlinespace
All males & 290 & $\rho$ & 0.41 & 0.40 & 0.39 & 0.47 & 0.36 & 0.35 & 0.47 & 0.31 & 0.29 \\
          &     & $r$    & 0.41 & 0.40 & 0.39 & 0.49 & 0.39 & 0.38 & 0.50 & 0.34 & 0.32 \\
\addlinespace
\ \ Male $\times$ Asian      &  52 & $\rho$ & 0.55 & 0.47 & 0.45 & 0.38 & 0.34 & 0.33 & 0.55 & 0.44 & 0.39 \\
                             &     & $r$    & 0.52 & 0.45 & 0.45 & 0.39 & 0.37 & 0.35 & 0.55 & 0.46 & 0.42 \\
\addlinespace
\ \ Male $\times$ African    &  93 & $\rho$ & 0.38 & 0.29 & 0.31 & 0.44 & 0.27 & 0.29 & 0.38 & 0.23 & 0.22 \\
                             &     & $r$    & 0.41 & 0.33 & 0.33 & 0.45 & 0.30 & 0.31 & 0.43 & 0.27 & 0.26 \\
\addlinespace
\ \ Male $\times$ Hispanic   &  52 & $\rho$ & 0.36 & 0.37 & 0.31 & 0.53 & 0.41 & 0.34 & 0.57 & 0.40 & 0.34 \\
                             &     & $r$    & 0.40 & 0.36 & 0.30 & 0.51 & 0.39 & 0.33 & 0.51 & 0.35 & 0.30 \\
\addlinespace
\ \ Male $\times$ White      &  93 & $\rho$ & 0.43 & 0.41 & 0.39 & 0.55 & 0.50 & 0.48 & 0.60 & 0.48 & 0.49 \\
                             &     & $r$    & 0.45 & 0.42 & 0.40 & 0.58 & 0.53 & 0.51 & 0.64 & 0.52 & 0.52 \\
\bottomrule
\multicolumn{12}{p{2.6in}}{Notes: Ensemble based on DINOv3, Inception and EVA02.}\\
\end{tabular}

%% file: tables/tabl_cfd_means_v1.tex

\begin{tabular}{lrcccc}
\toprule
 & & & \multicolumn{3}{c}{RVE Ensemble (Euclidean distance)} \\
\cmidrule(lr){4-6}
Subgroup & $n$ & HRS & Same gender & Same race/gender & All celebrities \\
\midrule
All                          & 597 & 3.23 & 0.30 & 0.21 & 0.31 \\
\addlinespace
Female                       & 307 & 3.45 & 0.29 & 0.16 & 0.31 \\
\ \ Female $\times$ Asian    &  57 & 3.64 & 0.30 & 0.15 & 0.29 \\
\ \ Female $\times$ Black    & 104 & 3.33 & 0.26 & 0.16 & 0.31 \\
\ \ Female $\times$ Hispanic &  56 & 3.46 & 0.29 & 0.11 & 0.30 \\
\ \ Female $\times$ White    &  90 & 3.45 & 0.30 & 0.19 & 0.33 \\
\addlinespace
Male                         & 290 & 3.00 & 0.31 & 0.26 & 0.30 \\
\ \ Male $\times$ Asian      &  52 & 2.85 & 0.27 & 0.24 & 0.30 \\
\ \ Male $\times$ Black      &  93 & 3.17 & 0.31 & 0.32 & 0.28 \\
\ \ Male $\times$ Hispanic   &  52 & 2.94 & 0.31 & 0.23 & 0.30 \\
\ \ Male $\times$ White      &  93 & 2.96 & 0.32 & 0.22 & 0.32 \\
\bottomrule
\end{tabular}

%% file: tables/cfd_regressions/tab_hrs_rve_regression.tex
\begin{tabular}{lcccc}
  \toprule
        &   (1)    &   (2)    &   (3)    &   (4)    \\
  \midrule
    RVE Ensemble & 3.036*** & 4.073*** & 4.284*** & 4.233*** \\
        & (0.216) & (0.188) & (0.190) & (0.192) \\
    \multicolumn{5}{l}{\emph{Controls}} \\
    Gender & \xmrk & \chkmrk & \chkmrk & \chkmrk \\
    Race & \xmrk & \xmrk & \chkmrk & \chkmrk \\
    Gender $\times$ Race & \xmrk & \xmrk & \xmrk & \chkmrk \\
    \bottomrule
    \multicolumn{5}{p{4.2in}}{Notes: Dependent variable: CFD human-rated attractiveness (HRS). $N=597$. Robust standard errors (HC3) in parentheses. *~p\textless.1, **~p\textless.05, ***~p\textless.01.} \\
\end{tabular}

%% file: tables/tabl_sample_composition.tex
\begin{tabular}{lccccc}
\toprule
 & Black & Asian & Hispanic & White & Total \\
\midrule
Female & 0 &  45 & 10 & 110 & 165 \\
Male   & 5 & 133 & 58 & 269 & 465 \\
All    & 5 & 178 & 68 & 379 & 630 \\
\bottomrule
\end{tabular}

%% file: tables/tabl_rve_jm.tex
\begin{tabular}{lrcccc}
\toprule
Subgroup & $n$ & Mean & Min & Median & Max \\
\midrule
All                          & 630 & 0.54 & 0.08 & 0.56 & 0.80 \\
\addlinespace
All Females                  & 165 & 0.57 & 0.15 & 0.57 & 0.80 \\
\ \ Female $\times$ Asian    &  45 & 0.53 & 0.19 & 0.56 & 0.70 \\
\ \ Female $\times$ Black    &   0 & \multicolumn{4}{c}{NA} \\
\ \ Female $\times$ Hispanic &  10 & 0.60 & 0.49 & 0.59 & 0.71 \\
\ \ Female $\times$ White    & 110 & 0.58 & 0.15 & 0.59 & 0.80 \\
\addlinespace
All Males                    & 465 & 0.53 & 0.08 & 0.56 & 0.79 \\
\ \ Male $\times$ Asian      & 133 & 0.47 & 0.08 & 0.48 & 0.76 \\
\ \ Male $\times$ Black      &   5 & 0.51 & 0.39 & 0.52 & 0.62 \\
\ \ Male $\times$ Hispanic   &  58 & 0.57 & 0.26 & 0.58 & 0.79 \\
\ \ Male $\times$ White      & 269 & 0.56 & 0.13 & 0.58 & 0.79 \\
\bottomrule
\multicolumn{6}{p{4.2in}}{Notes: Job market candidates sample. The RVE scores are calculated
using Euclidean distance and same-gender celebrities as a reference group.}\\
\end{tabular}

%% file: tables/tabl_outcome_defs.tex
\begin{tabular}{lp{1.85in}p{1.85in}c}
\toprule
Definition & Criteria & Brief Explanation & Success Rate \\
\midrule
BASELINE & Top-100 Academic OR Top-50 Policy & Standard measure of academic job market success & 27.6\% \\
\addlinespace
ALT.\ 1 & Same as Baseline MINUS Post-Docs \& Teaching & Excludes temporary and teaching-focused jobs & 18.1\% \\
\addlinespace
ALT.\ 2 & Baseline PLUS Top Tech Companies & Includes lucrative private sector as ``success'' & 29.8\% \\
\addlinespace
ALT.\ 3 & Top-150 Academic OR Top-75 Policy & Lower bar, more generous cutoff & 38.3\% \\
\addlinespace
ALT.\ 4 & Top-75 Academic OR Top-35 Policy & Higher bar, stricter cutoff & 23.7\% \\
\bottomrule
\end{tabular}

%% file: tables/tabl_LM_outcome1_RVE.tex
\begin{tabular}{lcccccc}
\toprule
       & \multicolumn{6}{c}{Outcome: Top-100 Academic or Top-50 Policy} \\
       & (1) & (2) & (3) & (4) & (5) & (6)  \\[1ex]
\midrule
RVE Ensemble & 0.431*** & 0.308** & 0.251*  & 0.185   & 0.185   & 0.152   \\
             & (0.127)  & (0.138) & (0.144)  & (0.142) & (0.141) & (0.141) \\[3ex]
RL Avg. Length    & &       &          & 0.280***&         & 0.233***\\
(,000 words)      & &       &          & (0.052) &         & (0.054) \\[3ex]
RL Avg. Sentiment & &       &          &         & 0.031***& 0.021***\\
(pp)              & &       &          &         & (0.007) & (0.007) \\[1ex]
\midrule
Adj.~$R^2$ & 0.014  & 0.043 & 0.072    & 0.119   & 0.101   & 0.130   \\
\multicolumn{7}{l}{\emph{Controls:}} \\
Candidate  & \xmrk & \chkmrk & \chkmrk & \chkmrk & \chkmrk & \chkmrk \\
PhD Program& \xmrk & \xmrk   & \chkmrk & \chkmrk & \chkmrk & \chkmrk \\
Year       & \xmrk & \xmrk   & \chkmrk & \chkmrk & \chkmrk & \chkmrk \\
RL Length  & \xmrk & \xmrk   & \xmrk   & \chkmrk & \xmrk   & \chkmrk \\
RL Sentiment& \xmrk & \xmrk  & \xmrk   & \xmrk   & \chkmrk & \chkmrk \\
\bottomrule
\multicolumn{7}{p{\textwidth}}{Notes: $N=630$. HC3 standard errors in parenthesis.}\\
\end{tabular} 

%% file: tables/tabl_LM_outcome1-5_RVE.tex
\begin{tabular}{lccccc}
\toprule
   & \multicolumn{5}{c}{Outcome} \\
   & Baseline & Alt. 1 & Alt. 2 & Alt. 3 & Alt. 4  \\[1ex]
\midrule
RVE Ensemble      & 0.152    & 0.137    & 0.122    & 0.226    & 0.091 \\
                  & (0.141)  & (0.125)  & (0.147)  & (0.156)  & (0.136) \\[3ex]
RL Avg. Length    & 0.233*** & 0.180*** & 0.216*** & 0.251*** & 0.178*** \\
(,000 words)      & (0.054)  & (0.052)  & (0.055)  & (0.057)  & (0.053) \\[3ex]
RL Avg. Sentiment & 0.021*** & 0.015**  & 0.023*** & 0.024*** & 0.022*** \\
(pp)              & (0.007)  & (0.006)  & (0.007)  & (0.008)  & (0.007) \\[1ex]
\midrule
Adj.~$R^2$        & 0.130    & 0.129    & 0.126    & 0.143    & 0.110\\
\multicolumn{6}{l}{\emph{Controls:}} \\
Candidate   & \chkmrk & \chkmrk & \chkmrk & \chkmrk & \chkmrk \\
PhD Program & \chkmrk & \chkmrk & \chkmrk & \chkmrk & \chkmrk \\
Year        & \chkmrk & \chkmrk & \chkmrk & \chkmrk & \chkmrk \\
RL Length   & \chkmrk & \chkmrk & \chkmrk & \chkmrk & \chkmrk \\
RL Sentiment& \chkmrk& \chkmrk & \chkmrk & \chkmrk & \chkmrk  \\
\bottomrule
\multicolumn{6}{p{\textwidth}}{Notes: $N=630$. HC3 standard errors in parenthesis.}\\
\end{tabular} 

%% file: tables/tabl_writers_desc_stats_presentation.tex
\begin{tabular}{lccc}
\toprule
  & Mean & Min & Max \\
\midrule
Adviser & 0.310 & 0.0 & 1.0 \\
Male & 0.881 & 0.0 & 1.0 \\
Experience & 19.41 & 1.0 & 57.0 \\
Top~10 RePEc & 0.514 & 0.0 & 1.0 \\
Top~5 RePEc & 0.348 & 0.0 & 1.0 \\
Full~Professor & 0.610 & 0.0 & 1.0 \\
\bottomrule
\multicolumn{4}{l}{$N=1925$} \\
\end{tabular} 

%% file: tables/tabl_reg_RL_rve_presentation.tex
\begin{tabular}{lccccc}
\toprule
     & (1) & (2) & (3) & (4) & (5)  \\[1ex]
\midrule
\multicolumn{6}{l}{\emph{Outcome: RL Word Count}} \\
RVE Ensemble & 393.0***   & 265.2**     & 252.2** & 254.3** & 241.6**   \\
             & (108.6)    & (107.2)     & (103.5) & (100.4) & (100.1)   \\[3ex]
\midrule
\multicolumn{6}{l}{\emph{Outcome: RL Sentiment}} \\
RVE Ensemble & 0.030***  & 0.022***   & 0.020***    & 0.020***     & 0.019**  \\
             & (0.008)   & (0.008)    & (0.008)     & (0.008)& (0.008)  \\[1ex]
\midrule
\multicolumn{6}{l}{\emph{Controls:}} \\
Candidate    & \xmrk    & \chkmrk  & \chkmrk & \chkmrk & \chkmrk  \\
Year         & \xmrk    & \chkmrk  & \chkmrk & \chkmrk & \chkmrk  \\
PhD Program  & \xmrk    & \xmrk    & \chkmrk & \chkmrk & \chkmrk  \\
Writer Char. & \xmrk    & \xmrk    & \xmrk   & \chkmrk & \chkmrk  \\
Writer Rank  & \xmrk    & \xmrk    & \xmrk   & \xmrk   & \chkmrk  \\
\bottomrule
\multicolumn{6}{p{3.2in}}{Notes: Euclidean metric and same-gender ref. group. $N=1,925$. Candidate-level clustered standard errors in parenthesis.}\\
\end{tabular} 

%% file: tables/tabl_reg_RL_rve_gender_presentation.tex
\begin{tabular}{lcc}
\toprule
                         & \multicolumn{2}{c}{\emph{Outcome}} \\
                         & RL Length & RL Sentiment \\[1ex]
\midrule
Male                     & 126.9     &   0.011     \\
                         & (123.5)   &  (0.009)    \\[2ex]
RVE Ensemble             & 399.9**   &   0.044***  \\
                         & (191.5)   &  (0.013)    \\[2ex]
RVE Ensemble$\times$Male & -212.5    &  -0.033**   \\
                         & (219.3)   &  (0.016)    \\[1ex]
\midrule
\multicolumn{3}{l}{\emph{Controls:}} \\
Candidate    & \chkmrk & \chkmrk  \\
Year         & \chkmrk & \chkmrk  \\
PhD Program  & \chkmrk & \chkmrk  \\
Writer Char. & \chkmrk & \chkmrk  \\
Writer Rank  & \chkmrk & \chkmrk  \\
\bottomrule
\multicolumn{3}{p{2.2in}}{Notes: Euclidean metric and same-gender ref. group. $N=1,925$. Candidate-level clustered standard errors in parenthesis.}\\

\end{tabular}